\documentclass{article}
\usepackage{spconf,amsmath,graphicx,hyperref, amssymb}
\usepackage{amsthm}
\usepackage{cite}
\usepackage{bm}
\usepackage{subcaption} 
\usepackage{bbm}
\usepackage{xcolor}
\usepackage{booktabs}
\usepackage{bm}
\usepackage{bbm}
\usepackage{cuted}
\usepackage{tikz}
\tikzset{every node/.style={font=\LARGE}}
\input{mySymbol.sty}

\usepackage{xcolor}

\newtheorem{theorem}{Theorem}

\theoremstyle{definition}
\newtheorem{definition}{Definition}[section]
\newtheorem{assumption}{Assumption}[section]
\newtheorem{proposition}{Proposition}[section]

\title{Graph Neural Networks in Large Scale Wireless Communication Networks: Scalability Across Random Geometric Graphs}
\name{Romina Garcia Camargo$^*$, Zhiyang Wang$^\dagger$, Alejandro Ribeiro$^*$\thanks{Supported by NSF TILOS and ARL DCIST CRA.}}
\address{$^*$Department of Electrical and Systems Engineering, University of Pennsylvania, Philadelphia, USA\\
$^\dagger$ Halıcıoğlu Data Science Institute, University of California San Diego, La Jolla, USA}
\begin{document}
\ninept
\maketitle
\begin{abstract}
The growing complexity of wireless systems has accelerated the move from traditional methods to learning-based solutions. Graph Neural Networks (GNNs) are especially well-suited here, since wireless networks can be naturally represented as graphs. A key property of GNNs is transferability: models trained on one graph often generalize to much larger graphs with little performance loss. While empirical studies have shown that GNN-based wireless policies transfer effectively, existing theoretical guarantees do not capture this phenomenon. Most works focus on dense graphs where node degrees scale with network size—an assumption that fails in wireless systems. In this work, we provide a formal theoretical foundation for transferability on Random Geometric Graphs (RGGs), a sparse and widely used model of wireless networks. We further validate our results through numerical experiments on power allocation, a fundamental resource management task.
\end{abstract}
\section{Introduction}
\label{sec:intro}
The use of machine learning to construct efficient wireless resource allocations has become popular. Graph Neural Networks (GNNs) have emerged as  a particularly effective architecture. Communication networks can often be modeled as graphs, with devices represented as nodes and their interactions as edges. This natural correspondence makes GNNs a strong candidate for tackling complex tasks in wireless systems \cite{eisen20, wang22, lu2024graph,lee2022graph}.

GNNs consist of stacked layers, each combining a graph convolutional filter with a point-wise nonlinearity \cite{ruiz21gnns,bruna14gnns,gama19gnns,deferrard16gnns}. Their adoption in wireless communication is motivated by properties such as permutation equivariance and stability \cite{vgarcia21,gama19stability,gama20stability}. Permutation equivariance enables learning independently of node labeling, improving data efficiency. Stability ensures that perturbations in the input lead to controlled perturbations in the output. These properties support learning policies that generalize across diverse and dynamic wireless network configurations.

A phenomenon of particular importance in wireless communication is transferability: policies trained on networks of one size often generalize to much larger networks with minimal performance loss. This effect has been observed empirically \cite{eisen20, eisen20transferability, fernandez24, wang22} and is in fact not unique to communications, but a widely studied property of GNNs \cite{ruiz23, maskey2022transferabilitygraphneuralnetworks, keriven20, wang2023geometricgraphfiltersneural, levin25}. Existing theoretical works consider graphs in the limit, commonly via graphons \cite{ruiz23} or manifolds \cite{wang2023geometricgraphfiltersneural}, where node degrees grow with network size, yielding dense or relatively sparse graphs. Remarkably, these analyses do not extend to wireless systems, where node degrees remain bounded by physical constraints. In addition, most approaches rely on abstract random models and neglect the geometric structure intrinsic to communication networks. Addressing the gap forms the core contribution of our work.

We develop a theoretical framework for analyzing the transferability of GNN-based resource-allocation policies in wireless networks. The key idea is to relate Random Geometric Graphs (RGGs), which naturally model wireless topologies via random node placements with radius-based connectivity \cite{penrose2003random,haenggirgg09}, to Deterministic Grid Graphs (DGGs), whose regular structure admits a clear connection to the transferability exploited by convolutional neural networks \cite{owerko2023transferability}. Using DGGs as a surrogate “bridge,” we quantify the difference between an RGG and a DGG with same size and density through a simple measure of matrix difference. This pair lets us derive transfer guarantees across scales, that is, if the RGG–DGG difference is sufficiently small, then a policy learned at one scale transfers to other scales of RGGs with a provably bounded loss (Theorem \ref{thm:trans-loss}). In particular, the transfer loss grows at most linearly with the RGG–DGG discrepancy. We validate the theory on the classical power-allocation task, training a GNN policy at one network scale and evaluating it across varying sizes. This contributes to both the theory of GNNs over sparse graphs and the practical applications of GNNs.

\section{Resource Allocation with Graph Neural Networks}
\label{sec:pf}
We consider the problem of resource allocation in a wireless network with $n$ users. At each time slot $t$, user $i$ is associated with a state $[\bbx(t)]_i$ (e.g., queue length or priority), summarized in the vector $\mathbf{x}(t)\in\mathbb{R}^n$. The channel gain from user $i$ to user $j$ is denoted $s_{ij}(t)$, with all channel gains collected in $\mathbf{S}(t)\in\mathbb{R}^{n\times n}$. The resource allocation policy is represented by $\mathbf{p}\in\mathbb{R}^n$. 
At each time step, the controller observes the network states $(\mathbf{x}(t),\mathbf{S}(t))$, selects the allocation strategy $\mathbf{p}(t)$, and receives a reward determined by the system. We define the expected reward $\mathbf{f}(\mathbf{p}(t);\mathbf{x}(t),\mathbf{S}(t))$ as the system reward and focus on the long-term average performance:
\begin{align}
    \mathbf{r} = \mathbb{E}[\mathbf{f}
    (\mathbf{p}; \mathbf{x}, \mathbf{S})], \label{eq:reward}
\end{align}
where the expectation is taken over the stationary joint distribution of $(\mathbf{x},\mathbf{S})$. This captures user experience under fast time-varying channels and states. The goal is to design a policy $\mathbf{p}(\mathbf{x},\mathbf{S})$ that maximizes expected reward \eqref{eq:reward}. We introduce a utility function $u_0(\mathbf{r})$ to formulate the optimization problem:
\begin{align}
 \mathbf{p}^\star(\mathbf{S}, \mathbf{x}) = & \argmax_{\mathbf{p}(\mathbf{x},\mathbf{S}) \in \mathcal{P}(\mathbf{x},\mathbf{S})} \ 
    u_0(\mathbf{r}), \label{eq:pa}
    \\
    \text{s.t.} \ \mathbf{r} =& \mathbb{E}[\mathbf{f}(\mathbf{p}(\mathbf{x},\mathbf{S}); \mathbf{x},\mathbf{S})], \nonumber\\
  \mathbf{u}(\mathbf{r}) & \geq\mathbf{0}, \nonumber
\end{align}
where $\mathbf{u}(\cdot)$ captures long-term system constraints (e.g., power budgets). This formulation is challenging as the objective is often nonconvex in $\mathbf{p}$. To address this, we introduce a parameterized policy $\bm{\Phi}(\mathbf{x},\mathbf{S}; \mathbf{H})$,
with parameters $\mathbf{H}\in\mathbb{R}^{s}$. The problem becomes
\begin{align}
 \mathbf{H}^\star = & \argmax_{\mathbf{H}\in\mathbb{R}^s} \ 
    u_0(\mathbf{r}), \label{eq:pa-lp}\\
    \text{s.t.} \ \mathbf{r} =& \mathbb{E}[\bm{\Phi}(\mathbf{x},\mathbf{S};\mathbf{H}); \mathbf{x},\mathbf{S})], \nonumber\\
 \mathbf{u}(\mathbf{r}) &\geq \mathbf{0}, \nonumber
\end{align}
The focus thus shifts from computing allocations directly to learning the parameters $\mathbf{H}$ of a policy class $\bm{\Phi}$. In this work, $\bm{\Phi}$ will be instantiated as a Graph Neural Network, which we describe next. For more details on the problem formulation see \cite{eisen20}.
\subsection{Graph Neural Networks}

A graph convolutional filter is a polynomial on a matrix representation of the graph. Considering a graph signal $\mathbf{x}\in\mathbb{R}^n$ (i.e. a vector supported on the nodes of a graph), we define a graph filter of order $K$ as follows \cite{ruiz21gnns, 7926424, Sandryhaila_2013, scaresllignn}:
\begin{align}
\mathbf{y} = \sum_{k=0}^{K-1} h_{k} \mathbf{S}^k \mathbf{x},
\end{align}
where $\{h_k\}_{k=0}^{K-1}$ are the filter coefficients and $\mathbf{S}\in\mathbb{R}^{n\times n}$ is the matrix representation of the graph, commonly referred to as the graph shift operator (GSO) \cite{shuman2013emerging}. 
As the GSO is symmetric, it is possible to diagonalize it as $\mathbf{S} = \mathbf{V\Lambda V}^H$, with $\mathbf{V}$ a matrix with the eigenvectors and $\mathbf{\Lambda}$ a diagonal matrix with the eigenvalues, namely $\bm\lambda = [\lambda_1, \lambda_2, \cdots, \lambda_K]$. Through a change of basis it is possible to obtain the spectral representation of the graph convolution, also denoted the graph frequency response of the filter:
\begin{align}
\hat{h}(\lambda) = \sum_{k=0}^{K-1} h_{k} \lambda^k.
\end{align}

Graph Neural Networks (GNNs) are composed of multiple layers, each combining a graph convolutional filter with a point-wise nonlinearity $\sigma$, with $\sigma : \mathbb{R} \rightarrow \mathbb{R}$. At the $l$-th layer, the filter takes as input graph signal $\mathbf{x}_{l-1} \in \mathbb{R}^d_{l-1}$, the output of the previous layer. This signal is then passed through $\sigma$:
\begin{align}
\mathbf{x}_l = \sigma\left(\sum_{k=0}^{K-1} h_{lk} \mathbf{S}^k \mathbf{x}_l\right).
\end{align}
This process is repeated across $L$ layers. The full set of trainable parameters is denoted $\mathbf{H} \in \mathcal{H}$, comprising all $h_{lk}$ for $l \in {1, \dots, L}$ and $k \in {0, \dots, K-1}$. Importantly, the dimensionality of $\mathbf{H}$ does not depend on the size of the graph.

\subsection{Random Geometric Graphs}
\label{sec:rgg}
A deterministic grid graph (DGG) $\mathbf{G}_n = (\mathcal{D}_n, \mathcal{E})$ is a graph defined on a regular lattice in a Euclidean space. Each of the $n$ nodes corresponds to a lattice point, and edges connect nodes that are direct neighbors on the grid. 
A Random Geometric Graph (RGG) \cite{penrose2003random} is an undirected graph $G^r_n = (\mathcal{V}_n, \mathcal{E}_r)$ constructed by placing $n$ nodes uniformly at random in a metric space of size $L_e \times L_e$:
\begin{align}
    \mathcal{V}_n \sim \mathcal{U}^2(0, L_e).
\end{align}
This uniform placement captures the \textit{random} aspect. An edge $(i,j) \in \mathcal{E}_r$ is included whenever the Euclidean distance between nodes $i$ and $j$ is at most a fixed connection radius $r_c$:
\begin{align}
    \mathcal{E}_r = \{(i,j) : d(i,j) \leq r_c\},
\end{align}
which explains the \textit{geometric} aspect. Suppose the density of the agents over the space is fixed as $\rho$, the expected number of neighbors of each agent is $\pi r_c^2/\rho$, which is also the average vertex degree. RGGs naturally arise in wireless communication settings, where nodes represent users or devices and connectivity within a fixed radius approximates feasible links determined by signal strength and interference.

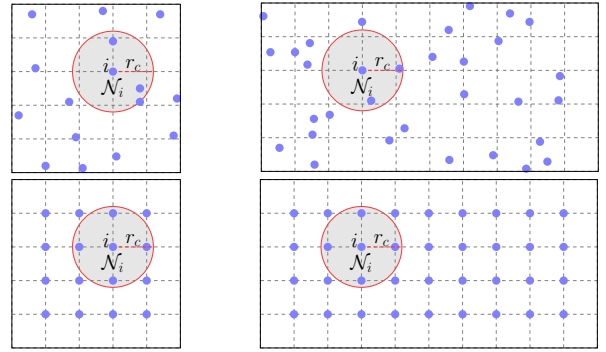
\begin{figure}[t]
\centering
\resizebox{0.45\textwidth}{!}{
  \begin{minipage}{\textwidth}
    \begin{subfigure}[b]{0.4\textwidth} 
        \usetikzlibrary{shapes.geometric, arrows}
\usetikzlibrary{arrows.meta}
\usetikzlibrary{positioning}
\usetikzlibrary{calc}
\usetikzlibrary{fit}
\usetikzlibrary{patterns}
\usetikzlibrary{decorations.markings}
\usetikzlibrary{backgrounds}

\begin{tikzpicture}
 \draw[color=red!80, fill = gray!20] (3,3) circle (1.2cm);
  \draw[help lines,dashed] (0,0) grid (5,5);

  \fill[color=blue!50] (canvas cs:x=1cm,y=0.2cm)    circle (3.5pt);
  \fill[color=blue!50] (canvas cs:x=1.9cm,y= 1.05cm) circle (3.5pt);
  \fill[color=blue!50] (canvas cs:x=3.1cm,y=0.48cm)    circle (3.5pt);
  \fill[color=blue!50] (canvas cs:x=4.8cm,y=1.1cm) circle (3.5pt);
  \fill[color=blue!50] (canvas cs:x=0.2cm,y=1.7cm)    circle (3.5pt);
  \fill[color=blue!50] (canvas cs:x=1.7cm,y=2.1cm) circle (3.5pt);
  \fill[color=blue!50] (canvas cs:x=3.8cm,y=2.1cm)    circle (3.5pt);
  \fill[color=blue!50] (canvas cs:x=3.8cm,y=2.5cm) circle (3.5pt);
  \fill[color=blue!50] (canvas cs:x=4.9cm,y=2.2cm)    circle (3.5pt);
  \fill[color=blue!50] (canvas cs:x=0.7cm,y=3.1cm) circle (3.5pt);
  
  \fill[color=blue!50] (canvas cs:x=2.1cm,y=3.8) circle (3.5pt);
  \fill[color=blue!50] (canvas cs:x=3cm,y=3.9cm)    circle (3.5pt);
  \fill[color=blue!50] (canvas cs:x=0.6cm,y=4.7cm) circle (3.5pt);
  \fill[color=blue!50] (canvas cs:x=2.7cm,y=4.8cm)    circle (3.5pt);
  \fill[color=blue!50] (canvas cs:x=4.4cm,y=4.7cm) circle (3.5pt);

   \draw[red!80, densely dashed, semithick] (3,3) -- (4.2,3);
   \fill[color=blue!50] (canvas cs:x=3cm,y=3cm)    circle (3.5pt);
   \node (A) at (2.8, 3.2)    {$i$};
   \node (B) at (3.6, 3.2)    {$r_c$};
   \node (C) at (3, 2.5)    {$\mathcal{N}_i$};
    \draw (0,0)     -- +(5,0) -- +(5,5) -- +(0,5) -- cycle;
\end{tikzpicture} 
        \label{fig:RGG}
    \end{subfigure}%
    \hfill
    \begin{subfigure}[b]{0.6\textwidth}
        \centering
        \usetikzlibrary{shapes.geometric, arrows}
\usetikzlibrary{arrows.meta}
\usetikzlibrary{positioning}
\usetikzlibrary{calc}
\usetikzlibrary{fit}
\usetikzlibrary{patterns}
\usetikzlibrary{decorations.markings}
\usetikzlibrary{backgrounds}

\begin{tikzpicture}
 \draw[color=red!80, fill = gray!20] (3,3) circle (1.2cm);
  \draw[help lines,dashed] (0,0) grid (10,5);

  \fill[color=blue!50] (canvas cs:x=5.56cm,y=4.2cm)    circle (3.5pt);
  \fill[color=blue!50] (canvas cs:x=5.08cm,y= 3.4cm) circle (3.5pt);
  \fill[color=blue!50] (canvas cs:x=6.88cm,y=0.48cm)    circle (3.5pt);
  \fill[color=blue!50] (canvas cs:x=4.25cm,y=1.28cm) circle (3.5pt);
  \fill[color=blue!50] (canvas cs:x=1.00cm,y=3.54cm)    circle (3.5pt);
  \fill[color=blue!50] (canvas cs:x=4.1cm,y=3.05cm) circle (3.5pt);
  \fill[color=blue!50] (canvas cs:x=8.86cm,y=2.83cm)    circle (3.5pt);
  \fill[color=blue!50] (canvas cs:x=0.54cm,y=0.70cm) circle (3.5pt);
  \fill[color=blue!50] (canvas cs:x=0.27cm,y=3.54cm)    circle (3.5pt);
  \fill[color=blue!50] (canvas cs:x=6.01cm,y=3.26cm) circle (3.5pt);
  \fill[color=blue!50] (canvas cs:x=7.86cm,y=2.27) circle (3.5pt);
  \fill[color=blue!50] (canvas cs:x=1.56cm,y=1.56cm)    circle (3.5pt);
  \fill[color=blue!50] (canvas cs:x=8.49cm,y=0.29cm) circle (3.5pt);
  \fill[color=blue!50] (canvas cs:x=7.63cm,y=2.07cm)    circle (3.5pt);
  \fill[color=blue!50] (canvas cs:x=6.19cm,y=4.91cm) circle (3.5pt);
  \fill[color=blue!50] (canvas cs:x=3.80cm,y=0.92cm)    circle (3.5pt);
  \fill[color=blue!50] (canvas cs:x=7.04cm,y=3.87cm) circle (3.5pt);
  \fill[color=blue!50] (canvas cs:x=1.44cm,y=3.81cm)    circle (3.5pt);
  \fill[color=blue!50] (canvas cs:x=1.58cm,y=0.20cm) circle (3.5pt);
  \fill[color=blue!50] (canvas cs:x=8.27cm,y=0.88 cm)    circle (3.5pt);
  \fill[color=blue!50] (canvas cs:x=2.99cm,y=4.44cm) circle (3.5pt);
  \fill[color=blue!50] (canvas cs:x=3.26cm,y=2.10 cm)    circle (3.5pt);
  \fill[color=blue!50] (canvas cs:x=6.02cm,y= 2.29cm) circle (3.5pt);
  \fill[color=blue!50] (canvas cs:x=1.52cm,y= 1.09cm)    circle (3.5pt);
  \fill[color=blue!50] (canvas cs:x=8.82cm,y= 2.11cm) circle (3.5pt);
  \fill[color=blue!50] (canvas cs:x=1.37cm,y= 3.17cm) circle (3.5pt);
  \fill[color=blue!50] (canvas cs:x=2.03cm,y= 1.68cm)    circle (3.5pt);
  \fill[color=blue!50] (canvas cs:x=7.38cm,y= 4.69cm) circle (3.5pt);
  \fill[color=blue!50] (canvas cs:x=7.96cm,y= 4.44cm)    circle (3.5pt);
  \fill[color=blue!50] (canvas cs:x=5.58cm,y= 0.34cm) circle (3.5pt);
   \fill[color=blue!50] (canvas cs:x=0.06cm,y= 4.61 cm)    circle (3.5pt);

   \draw[red!80, densely dashed, semithick] (3,3) -- (4.2,3);
   \fill[color=blue!50] (canvas cs:x=3cm,y=3cm)    circle (3.5pt);
   \node (A) at (2.8, 3.2)    {$i$};
   \node (B) at (3.6, 3.2)    {$r_c$};
   \node (C) at (3, 2.5)    {$\mathcal{N}_i$};
    \draw (0,0)     -- +(10,0) -- +(10,5) -- +(0,5) -- cycle;
\end{tikzpicture}
        \label{fig:RGG_large}
    \end{subfigure}

    \vspace{0.5em} 

    \begin{subfigure}[b]{0.4\textwidth} 
        \usetikzlibrary{shapes.geometric, arrows}
\usetikzlibrary{arrows.meta}
\usetikzlibrary{positioning}
\usetikzlibrary{calc}
\usetikzlibrary{fit}
\usetikzlibrary{patterns}
\usetikzlibrary{decorations.markings}
\usetikzlibrary{backgrounds}

\begin{tikzpicture}
 \draw[color=red!80, fill = gray!20] (3,3) circle (1.2cm);
  \draw[help lines,dashed] (0,0) grid (5,5);

  \fill[color=blue!50] (canvas cs:x=1cm,y=1cm)    circle (3.5pt);
  \fill[color=blue!50] (canvas cs:x=1 cm,y=2cm) circle (3.5pt);
  \fill[color=blue!50] (canvas cs:x=1cm,y=3cm)    circle (3.5pt);
  \fill[color=blue!50] (canvas cs:x=1cm,y=4cm) circle (3.5pt);
  \fill[color=blue!50] (canvas cs:x=2cm,y=1cm)    circle (3.5pt);
  \fill[color=blue!50] (canvas cs:x=2cm,y=2 cm) circle (3.5pt);
  \fill[color=blue!50] (canvas cs:x=2cm,y=3cm)    circle (3.5pt);
  \fill[color=blue!50] (canvas cs:x=2cm,y=4cm) circle (3.5pt);
  \fill[color=blue!50] (canvas cs:x=3cm,y=1cm)    circle (3.5pt);
  \fill[color=blue!50] (canvas cs:x=3cm,y=2cm) circle (3.5pt);
  
  \fill[color=blue!50] (canvas cs:x=4cm,y=4cm ) circle (3.5pt);
  \fill[color=blue!50] (canvas cs:x=3cm,y=4cm)    circle (3.5pt);
  \fill[color=blue!50] (canvas cs:x=4cm,y=1cm) circle (3.5pt);
  \fill[color=blue!50] (canvas cs:x=4cm,y=2cm)    circle (3.5pt);
  \fill[color=blue!50] (canvas cs:x=4cm,y=3cm) circle (3.5pt);

   \draw[red!80, densely dashed, semithick] (3,3) -- (4.2,3);
   \fill[color=blue!50] (canvas cs:x=3cm,y=3cm)    circle (3.5pt);
   \node (A) at (2.8, 3.2)    {$i$};
   \node (B) at (3.6, 3.2)    {$r_c$};
   \node (C) at (3, 2.5)    {$\mathcal{N}_i$};
    \draw (0,0)     -- +(5,0) -- +(5,5) -- +(0,5) -- cycle;
\end{tikzpicture} 
        \label{fig:DGG}
    \end{subfigure}%
    \hfill
    \begin{subfigure}[b]{0.6\textwidth}
        \centering
        \usetikzlibrary{shapes.geometric, arrows}
\usetikzlibrary{arrows.meta}
\usetikzlibrary{positioning}
\usetikzlibrary{calc}
\usetikzlibrary{fit}
\usetikzlibrary{patterns}
\usetikzlibrary{decorations.markings}
\usetikzlibrary{backgrounds}

\begin{tikzpicture}
 \draw[color=red!80, fill = gray!20] (3,3) circle (1.2cm);
  \draw[help lines,dashed] (0,0) grid (10,5);

  \fill[color=blue!50] (canvas cs:x=1cm,y=1cm)    circle (3.5pt);
  \fill[color=blue!50] (canvas cs:x=1 cm,y=2cm) circle (3.5pt);
  \fill[color=blue!50] (canvas cs:x=1cm,y=3cm)    circle (3.5pt);
  \fill[color=blue!50] (canvas cs:x=1cm,y=4cm) circle (3.5pt);
  \fill[color=blue!50] (canvas cs:x=2cm,y=1cm)    circle (3.5pt);
  \fill[color=blue!50] (canvas cs:x=2cm,y=2 cm) circle (3.5pt);
  \fill[color=blue!50] (canvas cs:x=2cm,y=3cm)    circle (3.5pt);
  \fill[color=blue!50] (canvas cs:x=2cm,y=4cm) circle (3.5pt);
  \fill[color=blue!50] (canvas cs:x=3cm,y=1cm)    circle (3.5pt);
  \fill[color=blue!50] (canvas cs:x=3cm,y=2cm) circle (3.5pt);
  
  \fill[color=blue!50] (canvas cs:x=4cm,y=4cm ) circle (3.5pt);
  \fill[color=blue!50] (canvas cs:x=3cm,y=4cm)    circle (3.5pt);
  \fill[color=blue!50] (canvas cs:x=4cm,y=1cm) circle (3.5pt);
  \fill[color=blue!50] (canvas cs:x=4cm,y=2cm)    circle (3.5pt);
  \fill[color=blue!50] (canvas cs:x=4cm,y=3cm) circle (3.5pt);

   \fill[color=blue!50] (canvas cs:x=5cm,y=1cm)    circle (3.5pt);
  \fill[color=blue!50] (canvas cs:x=5cm,y=2 cm) circle (3.5pt);
  \fill[color=blue!50] (canvas cs:x=5cm,y=3cm)    circle (3.5pt);
  \fill[color=blue!50] (canvas cs:x=5cm,y=4cm) circle (3.5pt);

   \fill[color=blue!50] (canvas cs:x=6cm,y=1cm)    circle (3.5pt);
  \fill[color=blue!50] (canvas cs:x=6cm,y=2 cm) circle (3.5pt);
  \fill[color=blue!50] (canvas cs:x=6cm,y=3cm)    circle (3.5pt);
  \fill[color=blue!50] (canvas cs:x=6cm,y=4cm) circle (3.5pt);
 \fill[color=blue!50] (canvas cs:x=7cm,y=1cm)    circle (3.5pt);
  \fill[color=blue!50] (canvas cs:x=7cm,y=2 cm) circle (3.5pt);
  \fill[color=blue!50] (canvas cs:x=7cm,y=3cm)    circle (3.5pt);
  \fill[color=blue!50] (canvas cs:x=7cm,y=4cm) circle (3.5pt);
   \fill[color=blue!50] (canvas cs:x=8cm,y=1cm)    circle (3.5pt);
  \fill[color=blue!50] (canvas cs:x=8cm,y=2 cm) circle (3.5pt);
  \fill[color=blue!50] (canvas cs:x=8cm,y=3cm)    circle (3.5pt);
  \fill[color=blue!50] (canvas cs:x=8cm,y=4cm) circle (3.5pt);
   \fill[color=blue!50] (canvas cs:x=9cm,y=1cm)    circle (3.5pt);
  \fill[color=blue!50] (canvas cs:x=9cm,y=2 cm) circle (3.5pt);
  \fill[color=blue!50] (canvas cs:x=9cm,y=3cm)    circle (3.5pt);
  \fill[color=blue!50] (canvas cs:x=9cm,y=4cm) circle (3.5pt);

   \draw[red!80, densely dashed, semithick] (3,3) -- (4.2,3);
   \fill[color=blue!50] (canvas cs:x=3cm,y=3cm)    circle (3.5pt);
   \node (A) at (2.8, 3.2)    {$i$};
   \node (B) at (3.6, 3.2)    {$r_c$};
   \node (C) at (3, 2.5)    {$\mathcal{N}_i$};
    \draw (0,0)     -- +(10,0) -- +(10,5) -- +(0,5) -- cycle;
\end{tikzpicture}
        \label{fig:DGG_large}
    \end{subfigure}
  \end{minipage}
}
\caption{Visualization of random geometric graphs as perturbations of deterministic grid graphs. \textbf{Top:} Illustrations of a small RGG (left) and large RGG (right). \textbf{Bottom:} Illustrations of a small DGG (left) and a large DGG (right).}
\label{fig:rgg-dgg}
\end{figure}

Let $\mathbf{S}_{\mathcal{D}_n}, \mathbf{S}_n \in \mathbb{R}^{n \times n}$ denote the adjacency matrices of $\mathbf{G}_n$, $\mathbf{G}^r_n$ respectively. If the norm difference between these matrices is sufficiently small, a RGG can be viewed as a perturbation of a DGG, obtained by adding Gaussian noise $\eta \sim \mathcal{N}(0,\sigma^2)$ to the node positions. This comparison, illustrated in Figure \ref{fig:rgg-dgg} for a radius $r_c$, forms the basis of our theoretical framework.

We first show that GNNs can transfer across scales on grid graphs from $\mathbf{G}_n$ to $\mathbf{G}_m$ (Theorem \ref{prop:grid-trans}). Next, we prove that a GNN trained on an RGG $\mathbf{G}_n^r$ close enough to a DGG $\mathbf{G}$ transfers with little performance loss (Theorem \ref{cor:diff-GNN}). Reversing the perturbation then yields transferability from $\mathbf{G}_m$ to a larger RGG $\mathbf{G}_{m}^r$ (Theorem~\ref{thm:trans-loss}). 

\section{Transferability of Graph neural Networks in Wireless Communication Networks}
\label{sec:transf}
We begin by establishing transferability across scales for deterministic grid graphs. Consider two DGGs, $\mathbf{G}_n$ and $\mathbf{G}_m$, with normalized adjacency matrices $\mathbf{S}_{\mathcal{D}_n}$ and $\mathbf{S}_{\mathcal{D}_m}$, where $n<m$. The adjacency matrix of a regular grid graph is circulant, with identical non-zero entries. This structure makes it possible to reinterpret the graph convolutional operation on a grid graph as a standard 2-D convolution, provided the nodes are indexed by their 2-D coordinates. Suppose that $n = B \times B$ with $B\in \mathbb{N}^+$ \footnote{Here $n$ could be decomposed in a general form $n=P\times B$ with zero-padding, we use this squared form for the ease of presentation.}, we reform the state matrix $\bbx\in \reals^{n}$ as a 2-d discrete function as 
\begin{equation}
    \label{eqn:signal-reform}
    x_B(n_1, n_2) = [\bbx]_{n_1 + n_2 \times B} \in \reals,
\end{equation}
with $n_1, n_2 = 0,1,\cdots, B-1$ representing the 2-d coordinates of each grid node. The graph convolution operation as a process of aggregating information from neighbors, is actually the same operation with the entries of the mask matrix equal to the non-zero entries of the adjacency matrix $\bbS_{\ccalD_n}$. The size of this mask matrix is related to the degree of each node, i.e. decided by $r_c$. We denote this mask matrix as $\bbL\in \reals^{M\times M}$, with $\smash[b] M = \lceil \sqrt{\pi r_c^2/\rho +1}\rceil$, which can be defined based on $\bbS_{\ccalD_n}$.
When we see the graph filter operating on the grid graphs, we can see it as a 2-D convolution operation. With the one-step aggregation rewritten as  
\begin{align}
     x_{B,1}(n_1,n_2) &=  \bbL\otimes x_B(n_1,n_2) \\
     &=\sum_{k_1=0}^{M-1} \sum_{k_2=0}^{M-1}  \bbL(k_1, k_2) x_B(n_1 -k_1, n_2-k_2),
\end{align}
which also collects signals over the neighboring nodes. Analogously, the $k$-step aggregation 
\begin{align}
     x_{B,k}&(n_1,n_2)   =  \bbL\otimes x_{B,k-1}(n_1,n_2)=  (\bbL\otimes )^k x_B(n_1,n_2).
    \end{align}

The graph convolution operation can be recovered by scaling $x_{B,k}$ with $h_k$ and summing up all the aggregated information, which is written as
    \begin{align}
    \label{eqn:gridfilter}
     y_{\bbL,B} = \sum_{k=0}^{K} h_k (\bbL\otimes )^k x_B := \bbh_D(\bbL,x_B).
\end{align}

Followed by a point-wise nonlinearity, this could recover the parameterization of $\bm\Phi(\bbx, \bbS_{\ccalD_n}; \bbH)$ with $L$ layers. We assume that the input signal $\mathbf{x}$ and the output performance $\mathbf{r}$ in \eqref{eq:reward} is jointly stationary over the 2-D space. We use the evaluation metric as the performance difference between the performance achieved by the learned policy $\bbr_n$ and the optimal policy $\bbr^*_n$, defined as 
\begin{equation}
\label{eqn:loss-perform}
    \mathcal{L}_n = \frac{1}{n}\| \mathbf{r}_n(\bm\Phi(\bbx, \bbS_{\ccalD_n}; \bbH)) - \mathbf{r}_n^*\|^2.
\end{equation}
\begin{theorem}
\label{prop:grid-trans}
Let 2-D convolutional neural network (i.e. GNN over a grid graph) be the parameterized policy that achieves a performance loss $\ccalL_n$ when applied on a grid graph with size $n$ and achieves a loss of $\ccalL_m$ when applied on another grid graph with size $m$. Suppose $n < m$, the difference of these two losses can be bounded as 
    \begin{align}
        \label{eqn:loss-diff-bound}
       &   \mathcal{L}_m \leq \mathcal{L}_n + C_M \mathbb{E}[\|\mathbf{x}\|^2]+\sqrt{\mathcal{L}_n C_M \mathbb{E}[\|\mathbf{x}\|^2] },
    \end{align}
    with the input and output signals are jointly stationary and bounded. $C_M =\frac{H_K^2}{n}[2\sqrt{n}KM+K^2M^2]$  with $H_K =\sum_{l=0}^K\sum_{k=0}^{K}|h_{lk}|\|\mathbf{L}\|_1^k$. When the neural network is trained on the grid graph with size $n$, i.e. $\mathcal{L}_{n}\leq \epsilon$, the loss achieved by implementing the trained neural network on the grid graph with size $m$.
\end{theorem}

\begin{proof}
    See Appendix 1 in \cite{Fullversion}.
\end{proof}

Under this interpretation, the transferability of CNNs extends naturally to grid graphs with fixed density as the number of nodes increases. This observation provides the bridge to GNNs on RGGs if we establish the connection between GNNs on grid graphs and on RGGs with the same scale and density.

To study transferability for a GNN trained on an RGG $\mathbf{G}_n^r$ with normalized adjacency matrix $\mathbf{S}_n$ to the grid graph $\mathbf{G}_n$, we introduce the following definition and assumption.

\begin{definition} (Integral Lipschitz continuous filter)
\label{def:lipschitz}
A filter $\hat{h}$ is integral Lipschitz continuous with constant $C$ if its frequency response satisfies
\begin{equation}\label{eqn:filter_function_integral}
    |\hat{h}(a)-\hat{h}(b)|\leq \frac{C |a-b| }{(a+b)/2} \text{ for all } a,b \in (0,\infty).
\end{equation}
\end{definition}

\begin{assumption}(Normalized Lipschitz nonlinearity)\label{ass:activation}
 The nonlinearity $\sigma$ is normalized Lipschitz continuous, i.e., $|\sigma(a)-\sigma(b)|\leq |a-b|$, with $\sigma(0)=0$.
\end{assumption}
We note that this assumption is reasonable, since most common nonlinearity functions are normalized Lipschitz. We assume that the graph filters used in the GNN are integral Lipschitz continuous as defined in Definition \ref{def:lipschitz}. Furthermore, we assume that the difference between the RGG and DGG matrices $\mathbf{S}_n-\mathbf{S}_{\mathcal{D}_n}$ is small. 
We define the performance metric on RGGs similar to \eqref{eqn:loss-perform} as 
\begin{equation}
    \mathcal{L}_n^r = \frac{1}{n}\|\mathbf{r}(\bm{\Phi}(\mathbf{x}_{n}, \mathbf{S}_{n}; \mathbf{H})) - \mathbf{r}_n^{r*}\|^2,
\end{equation}
which is the comparison between the performance on the learned policy $\bm{\Phi}(\mathbf{x}_{n}, \mathbf{S}_{n}; \mathbf{H})$ and the optimal performance.
We can now conclude the transferability of GNNs from RGG to DGG in the form of Theorem \ref{cor:diff-GNN}.

\begin{theorem}
\label{cor:diff-GNN}
Let $\bm{\Phi}(\mathbf{x},\mathbf{S};\mathbf{H})$ be an 1-layer GNN applied on a random geometric graph $\mathbf{G}^r_n$ and a grid graph $\mathbf{G}_n$. We define $\mathcal{W}_n = \mathbf{S}_n-\mathbf{S}_{\mathcal{D}_n}$ such that $\mathbb{E}[\|\mathcal{W}_n^2\|] =O(n^{-\alpha})$ with $\alpha>0$. Suppose that the GNN is trained on $\bbG_n^r$ with $\ccalL^r_n \leq \epsilon$, The difference of the outputs of GNN with input graph signal $\mathbf{x}\in\mathbb{R}^{n}$ can be bounded as 
 \begin{align}
        \label{eqn:diff-loss}
    \left|  \ccalL_n  - \ccalL_n^r \right| \leq
 C^2n^{1-\alpha}\|\bbx\|^2 +2\sqrt{\epsilon} C n^{\frac{1-\alpha}{2}}\|\bbx\|.
    \end{align}
\end{theorem}
\begin{proof}
    See Appendix 2 in \cite{Fullversion}.
\end{proof}
\noindent This proves that the difference of the performances of a GNN on a DGG $\mathbf{G}_n$ and on a RGG $\mathbf{G}_n^r$ can be bounded. 

We have shown that GNNs on RGGs can transfer to DGGs with the same number of nodes when the adjacency matrices are close enough in Theorem \ref{cor:diff-GNN}. Theorem \ref{prop:grid-trans} further proves that GNNs transfer on DGGs with different number of nodes. The transference of GNNs on RGGs with different number of nodes can therefore be derived based on the triangle inequality.

\begin{theorem}
    \label{thm:trans-loss}
Let $\bm{\Phi}(\mathbf{x},\mathbf{S};\mathbf{H})$ be a $L$-layer GNN applied on a graph with GSO $\mathbf{S}$ and input $\mathbf{x}$. Suppose there are two random geometric graphs $\mathbf{G}^r_n$ with adjacency matrix $\mathbf{S}_{n}$ and $\mathbf{G}^r_m$ with adjacency matrix $\mathbf{S}_{m}$, such that $n<m$. The network $\bm{\Phi}$ has been trained to minimize $\mathcal{L}_n^r \leq \epsilon$. We take $\alpha = 2$ and omit the terms that have order smaller than $\sqrt{\epsilon}$,
\begin{align}
&\nonumber |\mathcal{L}_n^r -\mathcal{L}_m^r |=\\ 
    & \mathcal{O}\Bigg(\sqrt{\epsilon}\left(n^{-1/2}\|\mathbf{x}_{n}\| + m^{-1/2}\|\mathbf{x}_{m}\|\right) + n^{-1}\|\mathbf{x}_{n}\|^2  + m^{-1}\|\mathbf{x}_{m}\|^2 \Bigg).
\end{align}
\end{theorem}
 \begin{proof}
     See Appendix 3 in \cite{Fullversion}.
 \end{proof}

We can see from the theorem that a GNN trained on a small RGG (a small wireless network) can be transferred to a larger RGG (a larger wireless network) with the trained policy approximating the optimal policy well enough. The difference between these two performances decreases with the size of these two networks and depends on the spectral continuity of the filter functions in the GNN. This attests that the trained policy over a wireless network modeled as a random geometric graph, i.e. graph with limited degree, can be transferred across scales without retraining. This fills the theoretical gap of analyzing the transferability of GNNs across sparse random geometric graphs. We verify this conclusion in a real-world power allocation scenario in the following.

\section{Numerical Experiments}
\label{sec:results}
We present numerical simulations for the power allocation problem that support our theoretical results. The utility function $u_0$ is defined as the sum rate, where the rates $r$ are as follows: 
\begin{align}
    r_i:= \log \Big(1+\frac{|h_{ij}|^2\mathbf{p}_i(\mathbf{x, H})}{\eta^2+\sum_{k\neq i}|h_{kj}|^2\mathbf{p}_k(\mathbf{x, H})}\Big). \label{eq:rates}
\end{align}
We consider the capacity that each transmitter experiences over the noise $\eta^2$ introduced in the AWGN channel and the interference caused by other users. We seek to maximize the expectation of the sum capacity over channel realizations. Assuming a power budget $P_{max}$, we formulate a simplified version of \eqref{eq:pa}.
\begin{align}
 \mathbf{p}^*(\mathbf{x, H})=&   \argmax\limits_{\substack{
                \mathbf{p}(\mathbf{x, H})\in\{0,p_0\}^n
}}  \quad  \sum_{i=1}^n r_i\label{eq:pa-specific}\\
&\text{s.t.}  \quad \mathbb{E}[\mathbf{1}^\mathsf{T}\mathbf{p}(\mathbf{x, H})] \leq P_{max} \nonumber\\
& \mathbf{p}(\mathbf{x, H})\in \{0, p_0\}^n.\nonumber
\end{align}
The solution to \eqref{eq:pa-specific} can be obtained by defining a learning parameterization and operating in the Lagrangian dual domain to obtain the optimal policy. The approach is analogous to the one seen in \cite{eisen20}. The key difference lies in the construction of the channel matrix, which is aligned with the structure of a weighted adjacency matrix.

We construct RGGs for training datasets as perturbations of DGGs, as described in Section \ref{sec:rgg}. 
Two-dimensional Gaussian noise is added to perturb the grid graphs, obtaining RGGs representative of communication networks. Isolated nodes were removed from the network, which resulted in uneven average number of nodes in each dataset. The channel matrix was created considering a path loss coefficient and a fading channel gain. Different scales were considered to evaluate transferability. Each dataset consists of 100 graphs with an average link count of $n_k \simeq 500 + 100k$, for $k =\{0, 1, \dots, 7\}$. More details from the implementation, such as the architecture and hyperparameters, can be seen in the repository of the project.\footnote{The code implementation can be found in \url{https://github.com/romm32/rgg_transferability}}  
We sample power assignments interpreting the output of the GNN as probability of assignment and sampling Bernoulli variables for the binary allocation. We consider a policy variation of the heuristic baseline \textit{WMMSE} \cite{wmmse}, using its outputs as probabilities to sample Bernoulli trials.
The results for the evaluation of our algorithm on unseen data can be seen in Table \ref{tab:performance}. For the power constraint, we compute $\mathbf{1}^\mathsf{T}\mathbf{p}(\mathbf{x, H})-P_{max, k}$ for each dataset, dividing over $n_k$ to get average per-node power budget violation. It can be seen that our model outperforms the baseline, successfully finding optimal allocation policies.

\begin{table}[t]
    \centering
    \begin{tabular}{lcc}
      \toprule
       & Sum rate & Power constraint \\
       \midrule
      WMMSE   & $276.70\pm9.62$     & $(-1.26\pm1.94)\times10^{-2}$       \\
      GNN     & $771.91 \pm 6.63$     & $(-3.69\pm1.29)\times10^{-2}$         \\
      \bottomrule
    \end{tabular}
    \caption{Performance against WMMSE. We present mean and standard deviation across 10 experiments over 100 unseen graphs.}
    \label{tab:performance}
\end{table}

\begin{figure}[t]
    \centering
    \includegraphics[width=0.95\linewidth]{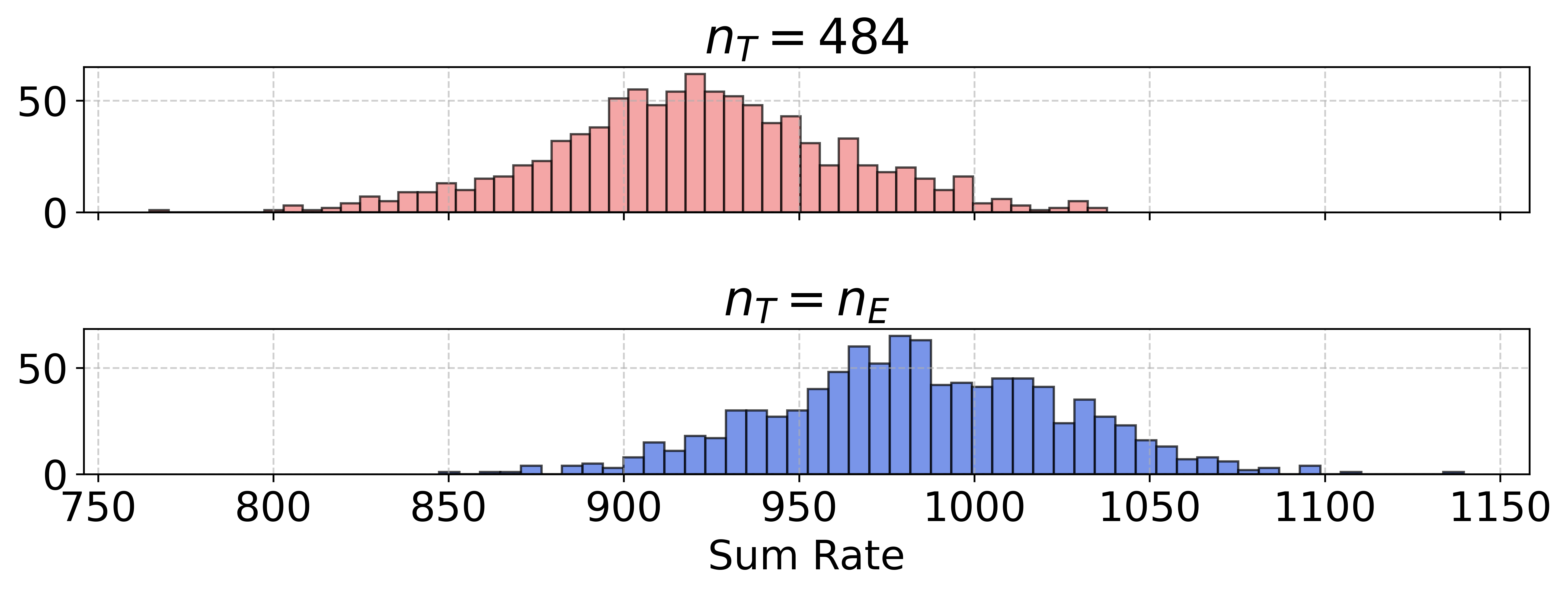}%
    \hfill
    \caption{Empirical distribution of sum rate achieved for 10 experiments. Comparison of a GNN trained for $n\simeq500$ and a GNN trained for $n\simeq 600$, both evaluated on a test dataset with $n\simeq600$.}
    \label{fig:histograms}
\end{figure}

\begin{figure}[h]
    \centering
    \includegraphics[width=0.95\linewidth]{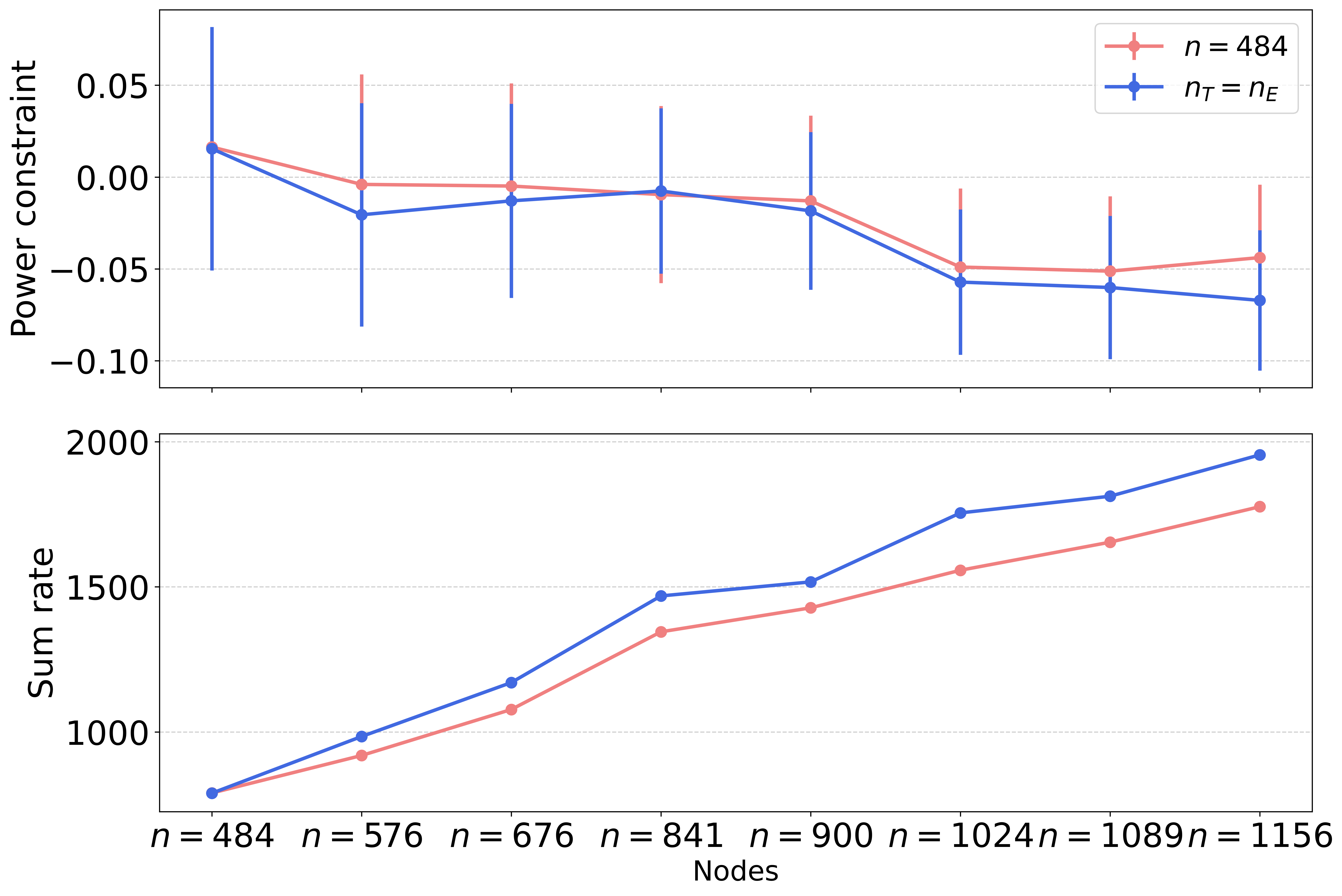}%
    \hfill
    \caption{Sum rate and power constraint values for networks of different scales with models trained in-distribution ($n_{T}=n_{E}$) and with a transferred model trained for $n=484$.}
    \label{fig:transferability1}
\end{figure}

Figures \ref{fig:histograms} and \ref{fig:transferability1} show transferability results for a model trained with $n \simeq 500$. For comparison, we trained separate models on each scale to evaluate in-distribution performance. Results are reported on graphs unseen during training. The transferred model performs on par with scale-specific models while maintaining low constraint violations. Overall, we observe a favorable trade-off between achieving high rates and avoiding both over- and under-allocation.

\section{Concluding remarks}
\label{sec:conc}
We presented a theoretical analysis of the transferability of wireless resource allocation policies using graph neural networks on random geometric graphs, and supported the findings with numerical experiments. Future work includes developing a more rigorous theoretical framework. It would also be valuable to investigate how transferability degrades when the underlying assumptions are not satisfied.




\appendix 
 {\section{Appendix}
 \subsection{Proof of Theorem \ref{prop:grid-trans}}
\subsubsection{Transferability of Filters between Grid Graphs }
Suppose there is a discrete random stationary signal over the 2-d space $f: \mathbb{N}\rightarrow \reals$, the input signal is a narrow window of this signal as 
$x_B= \sqcap_B f$, where $\sqcap_B (n_1,n_2) = \mathbbm{1}(0\leq n_1\leq B, 0\leq n_2\leq B)$. Next, we propose the transference of CNNs over different scales of grid graphs.
\begin{proposition}
\label{prop:grid-trans-filter}
Let $\bbh_D(\cdot)$ be a 2-d convolutional filter as defined in \eqref{eqn:gridfilter}. The output difference of the filter on a grid graph with size $n = B_1\times B_1$ and another grid graph with $m= B_2\times B_2$ can be bounded as
    \begin{align}
       & \nonumber \mathbb{E}[\|\sqcap_{B_1} (\bbh_D(\bbL, x_{B_1}) - \bbh_D(\bbL, x_{B_2}))\|^2_2] \\
        &\qquad \qquad \qquad \qquad \qquad \leq C_K^2 (B_2^2-B_1^2) \mathbb{E}[f(0,0)^2],
    \end{align}
    with $C_K = \sum\limits_{k=1}^K  \|\bbL\|_1^k|h_k|$.
\end{proposition}
\begin{proof}
    According to the definition of 2-d convolutional filter in \eqref{eqn:gridfilter}, the difference can be written as
\begin{align}
    & \nonumber \|\sqcap_{B_1} (\bbh_D(\bbL, x_{B_1}) - \bbh_D(\bbL, x_{B_2}))\| \\
    & = \|\sqcap_{B_1} (\bbh_D(\bbL, \sqcap_{B_1} f) - \bbh_D(\bbL, \sqcap_{B_2}f))\| \\
& = \Bigg\| \sqcap_{B_1}\left( \sum_{k=0}^K h_k (\bbL \otimes)^k (\sqcap_{B_1} f)  \right) - \\
&\qquad \qquad \qquad \sqcap_{B_1}\left( \sum_{k=0}^K h_k(\bbL \otimes)^k (\sqcap_{B_2} f) \right) \Bigg\|
\end{align}
With triangle inequality, we have 
\begin{align}
    & \nonumber \|\sqcap_{B_1} (\bbh_D(\bbL, x_{B_1}) - \bbh_D(\bbL, x_{B_2}))\| \\
    & \leq \sum_{k=0}^K \| \sqcap_{B_1} h_k (\bbL \otimes)^k (\sqcap_{B_1} f) - \sqcap_{B_1} h_k(\bbL \otimes)^k (\sqcap_{B_2} f) \|\\
    &\leq  \sum_{k=0}^K \|h_k(\bbL \otimes)^k (\sqcap_{B_1} f)   -  h_k(\bbL \otimes)^k (\sqcap_{B_2} f)  \|.
\end{align}
With Young’s convolution inequality 
\begin{equation}
   \| h_k \bbL \otimes (\sqcap_{B_1} f) \| \leq \| \bbL\|_1 \|h_k(\sqcap_{B_1} f)\|,
\end{equation}
we have
\begin{align}
     & \nonumber \|\sqcap_{B_1} (\bbh_D(\bbL, x_{B_1}) - \bbh_D(\bbL, x_{B_2}))\|\\
     &  \leq \sum_{k=0}^K \|\bbL \|_1^k |h_k| \|(\sqcap_{B_1} f)  -    (\sqcap_{B_2} f)  \| \\
     & = C_K \|(\sqcap_{B_1} f)  -    (\sqcap_{B_2} f)  \|,
\end{align}
if $B_1+ MK\geq B_2$.
With expectation, we have 
\begin{align}
   &\nonumber \mathbb{E} [ \|\sqcap_{B_1} (\bbh_D(\bbL, x_{B_1}) - \bbh_D(\bbL, x_{B_2}))\|^2  ]\\
   &\leq C_K^2 \mathbb{E}[\|(\sqcap_{B_1} f)  -    (\sqcap_{B_2} f)  \|^2 ]\\
& \leq C_K^2 (B_2^2 -B_1^2)\mathbb{E}[f(0,0)^2].
\end{align}
\end{proof}
\subsubsection{Transferability of GNNs between Grid Graphs }
We assume that the input signal $f$ and output $g$ are jointly stationary over the 2d space. The evaluation metric is a loss function of a supervised learning similar to the definitions in \cite{owerko2023transferability}. To simplify the notation, we assume that $n = B_1^2$ and $m = B_2 ^2$. To simplify the notation, we denote $\mathcal{L}_n= \ccalL_{B_1}$ and rewrite the output $\bm\Phi(\bbx, \bbS_{\ccalD_n};\bbH)$ as a 2-D signal $y_{\bbL,B_1}$ similar to \eqref{eqn:signal-reform}, specifically $y_{\bbL,B_1} = \sum_{k=0}^{K} h_k(\bbL\otimes )^k (\sqcap_{B_1} f)$ with a windowed input of $f$ which only inputs the values $f(i,j)$ if $i,j=0,1,\cdots, B_1-1$.
\begin{align}
    &\ccalL_{B_1}(\bbL) = \frac{1}{B_1^2} \mathbb{E}\left[ \sum_{i,j=0}^{B_1-1}  |y_{\bbL,B_1}(i,j) - g(i,j)|^2\right]. \label{eqn:loss-discrete}
\end{align}

We can conclude that the neural networks trained on a small-size grid graph can transfer to a larger grid graph with a bounded loss function.
\begin{proof}
    We denote the difference between the predicted outputs as $\epsilon(i,j)= y_\bbL(i,j)- g(i,j)$ with $y_\bbL$ as the output of the 2d-CNN when inputting $f$. For any time length $T$, let $N = \lceil B_2/B_1 \rceil$, then we have
    \begin{align}
      \ccalL_{B_2}( \bbL) \leq \mathbb{E}\left[ \frac{1}{(NB_1)^2}\sum_{n_1=0}^{NB_1-1} \sum_{n_2=0}^{NB_1-1} |\epsilon(i,j)|^2\right] . 
    \end{align}
    Recenter the summations and denote $\ccalT = \{ mB_1-\frac{(N-1)B_1}{2}| m\in \mathbb{Z}, 1\leq m<N\}^2$ as the center points, the summation can be decomposed as
    \begin{align}
  \ccalL_{B_2}( \bbL) & \leq \mathbb{E}\left[ \frac{1}{(NB_1)^2} \sum_{\bm\tau\in\ccalT} \left[ \sum_{i,j=0}^{B_1-1}   |\epsilon(i-\tau_1,j-\tau_2)|^2\right] \right]    \\
   &\leq  \frac{1}{(NB_1)^2} \sum_{\bm\tau\in\ccalT} \mathbb{E}\left[ \sum_{i,j=0}^{B_1-1} |\epsilon(i-\tau_1,j-\tau_2)|^2 \right].
    \end{align}
    With the inputs and outputs both stationary, we have $\mathbb{E}[|\epsilon(i-\tau_1, j-\tau_2)|^2] = \mathbb{E}[|\epsilon(i,j)|^2]$, which leads to
    \begin{align}
       &\ccalL_{B_2}( \bbL)\leq   \frac{1}{(NB_1)^2} \sum_{\bm\tau\in\ccalT}\mathbb{E}\left[ \sum_{i,j=0}^{B_1-1}  |\epsilon(i,j)|^2 \right]\\
      & \leq \frac{1}{B_1^2} \mathbb{E}\left[ \sum_{i,j=0}^{B_1-1}  |\epsilon(i,j)|^2 \right].
    \end{align}
    Next we replace $\epsilon$ with the indicator function and have an intermediate term $y_{\bbL,B_1}$ as the output of 2D-CNN when inputting $\sqcap_{B_1} f$.
    \begin{align}
  & \ccalL_{B_2}(\bbL) \leq \frac{1}{B_1^2} \mathbb{E}[\| \sqcap_{B_1} y_\bbL -g\|^2]\\
   & =\frac{1}{B_1^2}  \mathbb{E}[\| \sqcap_{B_1} y_\bbL- \sqcap_{B_1} y_{\bbL,B_1} + \sqcap_{B_1} y_{\bbL,B_1} -g\|^2]\\
   & \nonumber \leq \frac{1}{B_1^2}\mathbb{E}[\|  \sqcap_{B_1} y_{\bbL,B_1} -g\|^2] + \frac{1}{B_1^2}\mathbb{E}[\|  \sqcap_{B_1}( y_{\bbL,B_1} - y_\bbL)\|^2] \\
   &\label{eqn:3-1}  + \frac{2}{{B_1}^2}\mathbb{E}[\|  \sqcap_{B_1}( y_{\bbL,{B_1}} - g)\| \|\sqcap_{B_1}( y_{\bbL,{B_1}} - y_\bbL)\|] 
    \end{align}
The first term in \eqref{eqn:3-1} is $\ccalL_{B_1}( \bbL)$. The second term can be bounded as follows.
\begin{align}
    &\nonumber \|\sqcap_{B_1}(y_{\bbL} - y_{\bbL,{B_1}})\|\\
    & = \left\|\sqcap_{B_1}\left(\sum_{k=0}^{K-1} h_k(\bbL \otimes)^k f -\sum_{k=0}^{K-1} h_k(L\otimes)^k f_{B_1}\right)\right\|\\
    & \nonumber \leq \left\|\sqcap_{B_1}(f-f_{B_1}) \right\| + \left\|\sqcap_{B_1}h_1(\bbL\otimes f-\bbL\otimes f_{B_1}) \right\|+\cdots \\
    &\qquad \qquad\quad  +\left\|\sqcap_{B_1}h_{K-1}((\bbL\otimes)^{K-1}f-(\bbL\otimes)^{K-1}f_{B_1}) \right\|\\
    & \nonumber \leq \left\|\sqcap_{B_1}(f-\sqcap_Af) \right\|+ |h_1| \|\bbL\|_1 \|\sqcap_{{B_1}+M} (f -\sqcap_{B_1} f)\|_2 \\&\qquad \qquad + |h_2| \|\bbL\|_1^2 \|\sqcap_{{B_1}+2M} (f -\sqcap_{B_1} f)\|_2 \cdots\\
    &\leq \sum_{k=0}^{K-1} |h_k| \|\bbL\|_1^k \|\sqcap_{{B_1}+kM} (f -\sqcap_{B_1} f)\|_2 \\
    &\leq \sum_{k=0}^{K-1} |h_k| \|\bbL\|_1^k \|\sqcap_{{B_1}+(K-1)M} (f -\sqcap_{B_1} f)\|_2 \\
    &= H_K \|\sqcap_{{B_1}+(K-1)M} (f -\sqcap_{B_1} f)\|_2,
\end{align}
with $H_K= \sum_{k=0}^{K-1}|h_k|\|\bbL\|_1^k$. Therefore, we have
\begin{align}
   & \nonumber \ccalL_{B_2}( \bbL) \leq \ccalL_{B_1}( \bbL) + \frac{H_K^2}{B_1^2} \mathbb{E}\left[ \| \sqcap_{B_1+(K-1)M} (f-\sqcap_{B_1} f)\|^2\right] \\&\qquad + \sqrt{\ccalL_{B_1}(\bbL)\frac{H_K^2}{{B_1}^2} \mathbb{E}\left[ \| \sqcap_{{B_1}+(K-1)M} (f-\sqcap_{B_1} f)\|^2\right] }
\end{align}
Since $f$ is stationary, the second term can be seen as the variance of $f$ with a volume $({B_1}+(K-1)M)^2 - {B_1}^2$. Finally, we can derive 
\begin{align}
       \ccalL_{B_2}(\bbL) \leq \ccalL_{B_1}(\bbL) + C_M \mathbb{E}[f^2]+\sqrt{\ccalL_{B_1}(\bbL)C_M\mathbb{E}[f^2] },
    \end{align}
    where $C_M =\frac{H_K^2}{B_1^2}[2B_1KM+K^2M^2]$ with $H_K =\sum_{k=0}^{K}|h_k|\|\bbL\|_1^k$.
As we have normalized nonlinearities and multiple layers can be seen as a recurrent operation, the conclusion in Theorem \ref{prop:grid-trans} can be recovered.
\end{proof}

\subsection{Proof of Theorem \ref{cor:diff-GNN}}
\subsubsection{Transferability of Graph Filters across RGGs}
\begin{proposition}\label{thm:trans-rgg-dgg}
    Let $\bbh(\cdot)$ be a graph convolutional filter with integral Lipschitz continuous frequency responses with $|\lambda h'(\lambda)|\leq C$. Let $\bbS_n$ and $\bbS_{\ccalD_n}$ denote the adjancecy matrices of a random geometric graph and a deterministic geometric graph over a unit space respectively with $\ccalW_n = \bbS_n-\bbS_{\ccalD_n}$. If it satisfies that $\mathbb{E}[\left\|\ccalW_n^2\right\|] =O(1/n^\alpha)$ with $\alpha>0$. We have the difference of the outputs of graph convolutional filters with a input graph signal $\bbx\in\reals^{n }$ bounded as
    \begin{align}
        \mathbb{E}\left[\|\bbh(\bbS_{n}, \bbx) - \bbh(\bbS_{\ccalD_n}, \bbx)\|^2\right] \leq C^2 n^{1-\alpha}\|\bbx\|^2.
    \end{align}
\end{proposition}

\begin{proof}
    We denote $\ccalA_n = \bbS_n$ and $\ccalC_n =\bbS_{\ccalD_n}$ with $\ccalA_n = \ccalC_n + \ccalW_n$. With $\bby_A  =\sum_{k=0}^K h_k \ccalA_n^k \bbx$ and $\bby_C = \sum_{k=0}^K h_k \ccalC_n^k \bbx$, we have 
    \begin{align}
        &\nonumber \mathbb{E}[\|\bby_A-\bby_C\|^2]\\ 
        &=\mathbb{E}[tr(\bby_A \bby_A^\mathsf{T} -\bby_C\bby_C^\mathsf{T})] + 2\mathbb{E}[tr(\bby_C \bby_C^\mathsf{T} -\bby_A\bby_C^\mathsf{T})]\\
        & \nonumber = \sum_{k=0}^K \sum_{l=0}^K h_k h_l \left(\mathbb{E}[tr(\ccalA_n^k \bbx\bbx^\mathsf{T} \ccalA_n^l)] - tr(\ccalC_n^k \bbx\bbx^\mathsf{T} \ccalC_n^l) \right) \\
        &\label{eqn:1}+ 2 \sum_{k=0}^K \sum_{l=0}^K h_k h_l (tr(\ccalC_n^k \bbx\bbx^\mathsf{T} \ccalC_n^l) - \mathbb{E}[tr(\ccalA_n^k \bbx\bbx^\mathsf{T} \ccalC_n^l)]).
    \end{align}

 We start with the first term in \eqref{eqn:1} which can be written as 
\begin{strip}
    \begin{align}
& \nonumber \sum_{k,l=0}^K   h_k h_l \left(\mathbb{E}[tr(\ccalA_n^k \bbx\bbx^\mathsf{T} \ccalA_n^l)] - tr(\ccalC_n^k \bbx\bbx^\mathsf{T} \ccalC_n^l) \right)   \\
& = \sum_{k,l=0}^K h_k h_l( \mathbb{E}[tr(\ccalC_n+\ccalW_n)^k \bbx\bbx^\mathsf{T} (\ccalC_n+\ccalW_n)^l]   - tr(\ccalC_n^k \bbx\bbx^\mathsf{T} \ccalC_n^l) )\\
    &  \nonumber  = \sum_{k,l=0}^K h_k h_l\Bigg( \mathbb{E} \Bigg[tr(\ccalC_n^k \bbx\bbx^\mathsf{T} \ccalC_n^l + ((\ccalC_n+\ccalW_n)^k \bbx\bbx^\mathsf{T}\ccalC_n^l   -\ccalC_n^k \bbx\bbx^\mathsf{T} \ccalC_n^l ) +  ( \ccalC_n^k \bbx\bbx^\mathsf{T}(\ccalC_n+\ccalW_n)^l  -\ccalC_n^k \bbx\bbx^\mathsf{T} \ccalC_n^l))   \\& \qquad \qquad \qquad \qquad \qquad \qquad+tr\left(\left(\sum_{r=1}^k \ccalC_n^{k-r} \ccalW_n \ccalC_n^{r-1}\right) \bbx\bbx^\mathsf{T} \left( \sum_{s=1}^l \ccalC_n^{s-1} \ccalW_n \ccalC_n^{l-s}\right)\right)\Bigg]  -tr(\ccalC_n^k \bbx\bbx^\mathsf{T} \ccalC_n^l)]\Bigg) \\
    &\nonumber = \sum_{k,l=0}^K h_kh_l \Bigg(\mathbb{E}[tr((\ccalC_n+\ccalW_n)^k \bbx\bbx^\mathsf{T}\ccalC_n^l + \ccalC_n^k \bbx\bbx^\mathsf{T}(\ccalC_n+\ccalW_n)^l)] -2tr(\ccalC_n^k \bbx\bbx^\mathsf{T} \ccalC_n^l)  +  \\& \label{eqn:1-1}\qquad \qquad \qquad  \qquad \qquad \qquad  \mathbb{E}\left[tr\left(\left(\sum_{r=1}^k \ccalC_n^{k-r} \ccalW_n \ccalC_n^{r-1}\right) \bbx\bbx^\mathsf{T} \left( \sum_{s=1}^l \ccalC_n^{s-1} \ccalW_n \ccalC_n^{l-s}\right)\right)\right]\Bigg) +  \sum_{k,l=0}^K h_kh_l \mathbb{E}[tr(\ccalR_{kl})],
\end{align}
 
\end{strip}
where $\ccalR_{kl}$ represents the sum of the remaining terms that include terms with higher order than $\ccalW_n^2$.

The second term in \eqref{eqn:1} can be written as
\begin{align}
    \nonumber &2 \sum_{k,l=0}^K   h_k h_l tr(\ccalC_n^k \bbx\bbx^\mathsf{T} \ccalC_n^l) - \mathbb{E}[tr(\ccalA_n^k \bbx\bbx^\mathsf{T} \ccalC_n^l)]\\
    & \label{eqn:1-2}=\sum_{k,l=0}^K h_kh_l \Bigg(2tr(\ccalC_n^k \bbx\bbx^\mathsf{T} \ccalC_n^l) - 2\mathbb{E}[tr((\ccalC_n+\ccalW_n)^k \bbx\bbx^\mathsf{T}\ccalC_n^l)]\Bigg). 
\end{align}
We notice that putting \eqref{eqn:1-1} and \eqref{eqn:1-2} together leads to the expression of $\mathbb{E}[\|\bby_A-\bby_C\|^2]$ as
\begin{align}
    &\nonumber \mathbb{E}[\|\bby_A-\bby_C\|^2] = \sum_{k,l=0}^K h_kh_l \\&\mathbb{E}\left[tr\left(\left(\sum_{r=1}^k \ccalC_n^{k-r} \ccalW_n \ccalC_n^{r-1}\right) \bbx\bbx^\mathsf{T} \left( \sum_{s=1}^k \ccalC_n^{s-1} \ccalW_n \ccalC_n^{l-s}\right)\right)\right].
\end{align}
By moving all the summations outside, we have 
\begin{align}
    &\nonumber \mathbb{E}[\|\bby_A-\bby_C\|^2] = \sum_{k,l=0}^K \sum_{r=1}^k \sum_{s=1}^l h_kh_l \\&\mathbb{E}\left[tr\left(\left(  \ccalC_n^{k-r} \ccalW_n \ccalC_n^{r-1}\right) \bbx\bbx^\mathsf{T} \left(   \ccalC_n^{s-1} \ccalW_n \ccalC_n^{l-s}\right)\right)\right].
\end{align}
With the trace cyclic property $tr(\ccalA \ccalB \ccalC) = tr(\ccalC\ccalA\ccalB) = tr(\ccalB\ccalC\ccalA) $, we have 
\begin{align}
    &\nonumber \mathbb{E}[\|\bby_A-\bby_C\|^2] = \sum_{k,l=0}^K \sum_{r=1}^k \sum_{s=1}^l h_kh_l \\&\mathbb{E}\left[tr\left(   \ccalW_n\ccalC_n^{l+k-r-s} \ccalW_n \ccalC_n^{r-1}  \bbx\bbx^\mathsf{T}     \ccalC_n^{s-1} \right) \right].
\end{align}
For positive semidefinite matrices, the trace is submultiplicative as $tr(\ccalA\ccalB)\leq tr(\ccalA)tr(\ccalB)$. Therefore, we have
\begin{align}
    &\nonumber \mathbb{E}[\|\bby_A-\bby_C\|^2]\\
    &= \sum_{k,l=0}^K \sum_{r=1}^k \sum_{s=1}^l h_kh_l  \mathbb{E}\left[tr \left(  \ccalW_n\ccalC_n^{l+k-r-s} \ccalW_n\right)tr\left( \ccalC_n^{r-1}  \bbx\bbx^\mathsf{T}     \ccalC_n^{s-1} \right) \right]\\
    &\label{eqn:2} \leq \sum_{k,l=0}^K \sum_{r=1}^k \sum_{s=1}^l h_kh_l  \mathbb{E}\left[tr \left(  \ccalW_n\ccalC_n^{l+k-r-s} \ccalW_n\right)\right]tr\left( \ccalC_n^{r-1}  \bbx\bbx^\mathsf{T}     \ccalC_n^{s-1} \right).
\end{align}
In \eqref{eqn:2}, the deterministic term $tr(\ccalC_n^{r-1} \bbx\bbx^\mathsf{T} \ccalC_n^{s-1})$ can be decomposed with the spectral representation of $\bbx$. With $\bbx$ decomposed with respect to the eigenbasis $\{\bbe_i\}_{i=1}^n$ of $\ccalC_n$, $\bbx = \sum_{i=1}^n \hat{x}_i \bbe_{in}$, with $\hat{x}_i = \langle \bbx,\bbe_i \rangle$, we have 
\begin{align}
   &\nonumber  tr(\ccalC_n^{r-1} \bbx\bbx^\mathsf{T} \ccalC_n^{s-1})\\
    &=tr\left(\ccalC_n^{r-1} \left(\sum_{i=1}^n \hat{x}_i \bbe_i\right)\left(\sum_{i=1}^n \hat{x}_i \bbe_i\right)^{\mathsf{T}}\ccalC_n^{s-1}\right)\\
    & = tr\left( \left( \sum_{i=1}^n \hat{x}_i \ccalC_n^{r-1}  \bbe_i\right) \left(\sum_{i=1}^n \hat{x}_i \ccalC_n^{s-1}\bbe_i\right)^{\mathsf{T}}\right)\\
    & = tr\left( \left( \sum_{i=1}^n \hat{x}_i \lambda_i^{r-1}  \bbe_i\right) \left(\sum_{i=1}^n \hat{x}_i \lambda_i^{s-1}\bbe_i\right)^{\mathsf{T}}\right),
\end{align}
as the eigenbasis are orthonormal, i.e. $tr(\bbe_i\bbe_i^{\mathsf{T}}) = 1$ and $\bbe_i\bbe_j^\mathsf{T}=0$ for $i\neq j$, we have 
\begin{align}
    tr(\ccalC_n^{r-1} \bbx\bbx^\mathsf{T} \ccalC_n^{s-1}) &= tr\left(\sum_{i=1}^n \hat{x}_i^2 \lambda_i^{r+s-2} \bbe_i\bbe_i^{\mathsf{T}}\right)\\
    & = \sum_{i=1}^n \hat{x}_i^2 \lambda_i^{r+s-2}.
   \end{align}
Insert this back to \eqref{eqn:2}, we have
\begin{align}
   & \mathbb{E}[\|\bby_A-\bby_C\|^2]\\
    & \leq \sum_{i=1}^n \hat{x}_i^2 \sum_{k,l=0}^K \sum_{r=1}^k \sum_{s=1}^l h_kh_l  \mathbb{E}\left[tr \left(  \ccalW_n\ccalC_n^{l+k-r-s} \ccalW_n\right)\right]\lambda_i^{r+s-2}.
\end{align}
With the trace cyclic property and the inequality that 
\begin{equation}
    tr(\ccalA \ccalB)\leq \|\ccalA\|_2 tr(\ccalB),
\end{equation}
for any square matrix $\ccalA$ and positive semidefinite matrix $\ccalB$ \cite{wang1986trace}.
Therefore, the inequality can be derived further as
\begin{align}
& \mathbb{E}[\|\bby_A-\bby_C\|^2]\\
    & \leq \sum_{i=1}^n \hat{x}_i^2 \sum_{k,l=0}^K \sum_{r=1}^k \sum_{s=1}^l h_kh_l  \mathbb{E}\left[\|\ccalW_n^2\|_2\right] tr\left( \ccalC_n^{l+k-r-s}\right) \lambda_i^{r+s-2}
    \\& \leq \sum_{i=1}^n \hat{x}_i^2 \sum_{k,l=0}^K \sum_{r=1}^k \sum_{s=1}^l h_kh_l  \mathbb{E}\left[\|\ccalW_n^2\|_2\right] \sum_{j=1}^n \lambda_j^{l+k-r-s}  \lambda_i^{r+s-2}.
\end{align}
By changing the summation order, we have
\begin{align}
& \nonumber \mathbb{E}[\|\bby_A-\bby_C\|^2]\\
&\nonumber  \leq \sum_{i,j=0}^n\sum_{r,s=1}^K \left(\sum_{k=r}^K h_k\lambda_i^{r-1} \lambda_j^{k-r} \right)\left(\sum_{l=s}^K h_l \lambda_i^{s-1} \lambda_j^{l-s}\right) \\
&\qquad \qquad \qquad\qquad \qquad   \qquad \qquad\qquad \hat{x}_i^2 \mathbb{E}\left[\|\ccalW_n^2\|\right]\\
&\label{eqn:3}\leq \sum_{i,j=0}^n\sum_{r =1}^K\left(\sum_{k=r}^K h_k\lambda_i^{r-1} \lambda_j^{k-r} \right)^2\hat{x}_i^2 \mathbb{E}\left[\|\ccalW_n^2\|\right].
\end{align}
We import the definition of generalized Lipschitz gradient (Definition 4 in \cite{gao2021stability}) of generalized graph filter frequency response (Definition 3 in \cite{gao2021stability}). More specifically, let $\bm\lambda^{(r)} = [\lambda_i,\lambda_i\cdots,\lambda_j,\cdots,\lambda_j]^\mathsf{T}$ with $r$ $\lambda_i$ followed by $K-r$ $\lambda_j$ and $\bm\lambda^{(r)} \in \reals_+^{K}$.  The partial derivative of generalized $h(\bm\lambda)$ with respect to the $r$-th entry $\lambda_r$ as 
\begin{equation}
    \frac{\partial h(\bm\lambda^{(r)})}{\partial \lambda_r} = \sum_{k=r}^K h_k \lambda_i^{r-1} \lambda_j^{k-r}, \text{for all } r =1, 2\cdots,K.
\end{equation}
The generalized Lipshitz gradient between $\lambda_i$ and $\lambda_j$ is defined as 
\begin{equation}
    \nabla_L h(\lambda_i,\lambda_j) = \left[\frac{\partial h(\bm\lambda^{(1)})}{\partial \lambda_1},\cdots \frac{\partial h(\bm\lambda^{(K)})}{\partial \lambda_K}\right]^\mathsf{T}.
\end{equation}
Therefore, the inequality \eqref{eqn:3} can be further derived combined with the Lipschitz gradient assumption $\|  \nabla_L h(\lambda_i,\lambda_j)\|\leq C_L\leq C$ and $\|\mathbb{E}[\ccalW_n^2]\| = \ccalO(1/n^\alpha)$. 
\begin{align}
   & \mathbb{E}[\|\bby_A-\bby_C\|^2]\leq \sum_{i,j=0}^n \left\|\nabla_L h(\lambda_i,\lambda_j) \right\|^2\hat{x}_i^2 \mathbb{E}\left[\|\ccalW_n^2\|\right]\\
    &\leq n C^2 \mathbb{E}\left[\|\ccalW_n^2\|\right] \|\bbx\|^2\leq C^2 n^{1-\alpha} \|\bbx\|^2.
\end{align}
This concludes the proof.
\end{proof}

\subsubsection{Transferability of GNNs across RGGs}
\begin{proposition}
\label{prop:diff-GNN}
Let $\bm\Phi(\bbx,\bbS_n;\bbH)$ be an 1-layer GNN applied on a random geometric graph $\bbG^r_n$ and a grid graph $\bbG_n$. Under the same setting with Theorem \ref{thm:trans-rgg-dgg}, the difference of the outputs of GNN with input graph signal $\bbx\in\reals^{n}$ can be bounded as 
\begin{align}
    \label{eqn:diff-GNN}
    \mathbb{E}\!\left[\left\|\bm{\Phi}(\mathbf{x},\mathbf{S}_n;\mathbf{H}) 
    - \bm{\Phi}(\mathbf{x},\mathbf{S}_{\mathcal{D}_n};\mathbf{H})\right\|^2 \right]
    \leq F^L C^2 n^{1-\alpha}\|\mathbf{x}\|^2.
\end{align}
\end{proposition}
\begin{proof}
    To bound the output difference of GNNs on RGG and grid graph, we need to write in the form of features of the final layer
 \begin{align}
\nonumber &\mathbb{E}\left[\|\bm\Phi(\bbx;\bbS_n,\ccalH)- \bm\Phi (\bbx;\bbS_{\ccalD_n},\ccalH)\|^2\right] 
\\& = \sum_{q=1}^{F} \mathbb{E}\left[\left\| \sigma(\bby_{A,L}^q)- \sigma(\bby_{C,L}^q) \right\|^2\right]\\
& \leq \sum_{q=1}^{F} \mathbb{E}\left[\left\|  \bby_{A,L}^q -  \bby_{C,L}^q  \right\|^2\right],\\
&\leq F  \mathbb{E}\left[\left\|  \bby_{A,L-1}^q -  \bby_{C,L-1}^q  \right\|^2\right]
 \end{align}
 where the inequality comes from the normalized Lipschitz of nonlinearities. 
\end{proof}

\subsubsection{Proof of Theorem \ref{cor:diff-GNN}}
\begin{proof}
    We replace $\bby_A = \bm\Phi(\bbx, \bbS_n,\bbH)$ and $\bby_C = \bm\Phi(\bbx,\bbS_{\ccalD_n},\bbH)$ and $\bbr_n^* =\bbg$ for the ease of presentation. The MSE loss can be written as
    \begin{align}
    \nonumber   
    &\left|\mathbb{E}\left[\|\bby_A-\bbg\|^2 - \|\bby_C-\bbg\|^2 \right]\right|\\
    & \leq  \mathbb{E}\left[\left|\|\bby_A-\bbg\|^2 - \|\bby_C-\bbg\|^2\right| \right] \\
    &= \mathbb{E}\left[\|\bby_A-\bbg -\bby_C+\bbg\|\|\bby_A-\bbg + \bby_C-\bbg\| \right]\\
    &\leq \mathbb{E}\left[\|\bby_A -\bby_C\|(\|\bby_A-\bbg\| + \|\bby_C-\bbg\|) \right]\\
    &\leq \mathbb{E}\left[\|\bby_A -\bby_C\|(\|\bby_C-\bbg\| + \sqrt{\epsilon}) \right] \label{eqn:T5-1}
    \end{align}
By subtracting and adding $\bby_A$ in the term $\|\bby_C-\bbg\|$, we have 
\begin{equation}
    \|\bby_C-\bbg\|\leq \|\bby_A-\bby_C\|+\|\bby_A-\bbg\|,
\end{equation}
which depends on the triangle inequality. Inserting this into \eqref{eqn:T5-1}, we have
\begin{align}
\nonumber&\left|\mathbb{E}\left[\|\bby_A-\bbg\|^2 - \|\bby_C-\bbg\|^2 \right]\right|\\
    &\leq \mathbb{E}[\|\bby_A-\bby_C\|^2] +2\sqrt{\epsilon}\mathbb{E}[\|\bby_A-\bby_C\|].\label{eqn:T5-2}
\end{align}
With Jensen inequality, we have
\begin{equation}
    \mathbb{E}[\|\bby_A-\bby_C\|] \leq \sqrt{\mathbb{E}[\|\bby_A-\bby_C\|^2] }.
\end{equation}
With the conclusion in Proposition \ref{cor:diff-GNN}, we have
\begin{align}
\mathbb{E}\left[\left\|\bby_A - \bby_C\right\|^2 \right] \leq  C_L^2n^{1-\alpha}\|\bbx\|^2.
\end{align}
Bring this to \eqref{eqn:T5-2}, we have
\begin{align}
   \left|  \ccalL_n^r - \ccalL_n )   \right| \leq C^2n^{1-\alpha}\|\bbx\|^2 +2\sqrt{\epsilon} C n^{\frac{1-\alpha}{2}}\|\bbx\|, 
\end{align}
which concludes the proof.
\end{proof}

\subsection{Proof of Theorem \ref{thm:trans-loss}}
\begin{proof}
We first decompose the loss difference between GNN on $\bbS_{n}$ and $\bbS_{m}$ by inserting intermediate terms of loss of GNNs on $\bbS_{\ccalD_{n}}$ and $\bbS_{\ccalD_{m}}$.
    \begin{align}
       \label{eqn:4-1} |\ccalL_n^r- \ccalL_m^r| \leq | \ccalL_n^r - \ccalL_n| + |\ccalL_n-\ccalL_m| + |\ccalL_m -\ccalL_m^r|
    \end{align}
The first and the third terms in \eqref{eqn:4-1} can be bounded with Theorem \ref{prop:grid-trans}. The second term can be bounded with Theorem \ref{cor:diff-GNN}.
\end{proof}

}

\bibliographystyle{IEEEbib}
\bibliography{refs}

\end{document}